# First-principles study of the thermoelectric properties of Zintl compound KSnSb

S. Huang, H. J. Liu[*], D. D. Fan, P. H. Jiang, J. H. Liang, G. H. Cao, J. Shi

*Key Laboratory of Artificial Micro- and Nano-Structures of Ministry of Education and School of Physics and Technology, Wuhan University, Wuhan 430072, China*

The unique structure of Zintl phase makes it an ideal system to realize the concept of phonon-glass and electron-crystal in the thermoelectric community. In this work, by combining first-principles calculations and Boltzmann transport theory for both electrons and phonons, we demonstrate that the *ZT* value of Zintl compound KSnSb can reach ~2.6 at 800 K. Such extraordinary thermoelectric performance originates from the large Seebeck coefficient due to multi-valley band structures and particularly very small lattice thermal conductivity caused by mixed-bond characteristics.

## 1. Introduction

Energy is the backbone of modern civilization. Developing new and renewable energy has become an urgent issue. Among them, the thermoelectric materials can directly convert heat into electrical energy and *vice versa*, which turns out to be simple, durable and environmental-friend. The conversion efficiency of thermoelectric materials depends on the figure of merit (*ZT*) which can be expressed as [1]

$$ZT = \frac{S^2 \sigma T}{\kappa_e + \kappa_p}, \tag{1}$$

where $S$, $\sigma$, $T$, $\kappa_e$ and $\kappa_p$ are the Seebeck coefficient, the electrical conductivity, the absolute temperature, the electronic thermal conductivity, and the lattice thermal conductivity, respectively. In order to obtain higher *ZT* value, a thermoelectric material should have larger Seebeck coefficient, higher electrical conductivity and lower thermal conductivity. Slack proposed the concept of ideal thermoelectric material, namely, the phonon-glass and electron-crystal (PGEC) [1], which has very low thermal conductivity like phonon in glass, as well as high electrical conductivity like electron in crystal. It is believed that Zintl phases are typical representatives of PGEC materials and generally exhibit pretty good thermoelectric performance [2]. The complex structures of Zintl compounds, which are usually composed of the cations and the polyanions, contribute to the low thermal conductivity. In addition, the cations are often made up of group I and II elements which provide electrons to the polyanions, and the covalent bond in the polyanions can ensure relatively high carrier mobility. All these favorable characteristics could lead to good thermoelectric performance of Zintl compounds.

---

[*] Author to whom correspondence should be addressed. E-mail: phlhj@whu.edu.cn

In recent years, more and more Zintl compounds with higher thermoelectric performance have been found. For instance, Hu *et al.* reported that the Zintl material $Yb_{14}MgSb_{11}$ synthesized by annealing the mixture of elements can reach a high *ZT* value of 1.02 at 1075 K [3]. Yamada *et al.* prepared the polycrystalline sample of the Zintl phase $Na_{2+x}Ga_{2+x}Sn_{4-x}$ via a series of procedures including pulverization, mixing, compaction, and heating. They found that the *ZT* value is 1.28 at 340 K when $x = 0.19$ [4]. Shuai *et al.* synthesized Zintl compound $Eu_{0.2}Yb_{0.2}Ca_{0.6}Mg_2Bi_2$ by ball milling and hot pressing, and the *ZT* value can be as high as 1.3 at 873 K [5]. On the other hand, it is interesting to note that many Sb-based compounds were demonstrated to be good thermoelectric materials, such as $Sb_2Te_3$ [6], $\beta$-$Zn_4Sb_3$ [7], $Yb_{14}MnSb_{11}$ [8] and so on. It is thus natural to ask whether the Sb-based Zintl compound KSnSb [9] is a potential thermoelectric material. In fact, the *β* value [10], which is usually identified as a metric of thermoelectric performance, is calculated to be 79.6 for *n*-type KSnSb and even larger than that of the state-of-the-art thermoelectric material SnSe (22.9) [11]. Such observation indicates that KSnSb is likely to have good thermoelectric performance, which deserves deep investigation.

The KSnSb compound was firstly synthesized in 1987 and its nonmetallic behavior was suggested [12]. Subsequently, Schmidt *et al.* calculated its electronic band structure and density of states (DOS) by density functional theory (DFT). They adopted the concept of Zintl phase to understand the electronic properties of KSnSb, and found that the SnSb substructures exist stable and saturated covalent bonds because the free valence electrons of K atoms are transferred to the polyanion [13]. In a later work of Alemany *et al.*, first-principles calculations on the Zintl phase KSnSb was carried out and they clearly pointed out that the system should be semiconducting [14]. However, to the best of our knowledge, the thermoelectric properties of Zintl compound KSnSb are never mentioned before. In this work, by using first-principles calculations combined with Boltzmann transport theory, we give a detailed investigation of the electronic, phonon, and thermoelectric transport properties of KSnSb compound. We shall see that a record high *ZT* value of ~2.6 can be realized at 800 K by optimizing the carrier concentration, which suggests the intriguing thermoelectric application potentials of such Zintl compound.

## 2. Computational methods

The electronic properties of KSnSb are calculated by using the first-principles projector augmented-wave (PAW) method within the framework of DFT, which is implemented in the so-called Vienna *ab-initio* simulation package (VASP) [15−17]. The exchange and correlation energy is in the form of Perdew-Burke-Ernzerhof (PBE) [18]. To accurately predict the band gap, we also employ the hybrid density functional of Heyd–Scuseria–Ernzerhof (HSE06) [19-21]. The kinetic energy cutoff of the plane-wave basis is set to be 400 eV with a Γ-center $12\times12\times4$ *k*-mesh. During the structure relaxation, the energy convergence threshold is set to $10^{-6}$ eV. The van der Waals (vdW) corrections in the form of optB86b [22] are considered. Based on the band structure, the electronic transport coefficients (the Seebeck coefficient, the electrical conductivity, and the electronic thermal conductivity) can be readily calculated by using the Boltzmann transport theory [23] and the rigid band picture [24]. On the other hand, the lattice thermal conductivity can be obtained from the Boltzmann transport theory for phonons [25]. A $6\times6\times2$ supercell is adopted for the calculation of 2nd-order interatomic force constants (IFCs) [26]. For the 3rd-order IFCs [27], the maximum cutoff distance is set to be 8th neighbor in the supercell. The lattice thermal conductivity is calculated with a *q*-point grid of 24×24×8, and the scale parameter for smearing is set to 0.1.

## 3. Results and discussions
### 3.1 Electronic and transport properties

The KSnSb compound has a typical layered structure which is made up of the strong electropositive cation $K^+$ and the covalent bonding polyanion $(SnSb)^-$. There are ionic bonds and vdW interactions between K and SnSb layers. Figure 1 shows the crystal structure of KSnSb, which has the space group of *P*6$_3$*mc* [12]. The primitive cell is hexagonal and contains 6 atoms, where the K atoms occupy the site (2/3, 1/3, 0) and (1/3, 2/3, 1/2), the Sn atoms are located at (0, 0, $z_{Sn}$) and (0, 0, 1/2 + $z_{Sn}$), and the Sb atoms at (2/3, 1/3, $z_{Sb}$) and (1/3, 2/3, 1/2 + $z_{Sb}$). The internal coordinates $z_{Sn}$ and $z_{Sb}$ and the lattice constants of KSnSb are summarized in Table I, where we see that the results with vdW corrections give better agreement with the experimental data than those without vdW interactions.

Figure 2 plots the energy band structure of KSnSb, where the calculations using PBE and HSE06 functionals are both shown for comparison. In each case, we see that the conduction band minimum (CBM) is located at Γ point. Besides, there is a conduction band extreme (CBE) along the ΓM direction with energy slightly higher (~0.09 eV) than that of CBM. As for the valence band maximum (VBM), we observe two valleys with almost identical energy along the ΓM and ΓK directions. Such kind of multi-valley structures usually lead to large power factor ($PF=S^2\sigma$) [28], as previously found in the Bi$_2$Te$_3$, Sb$_2$Te$_3$ [29] and SnSe [30] systems. We will come back to this point later. The indirect band gap of KSnSb is calculated to be 0.31 eV using the PBE functional. It is well known that standard DFT tends to underestimate the band gap seriously. As an alternative, the HSE06 calculation predicts a larger gap of 0.75 eV which is exclusively used in the following calculations.

Within the framework of Boltzmann transport theory [23], the electrical conductivity $\sigma$, the Seebeck coefficient $S$, and the electronic thermal conductivity

$\kappa_e$ can be obtained from the band structure

$$S = \frac{ek_B}{\sigma} \int \Xi(\varepsilon)\left(-\frac{\partial f_0}{\partial \varepsilon}\right)\frac{\varepsilon - \mu}{k_B T} d\varepsilon, \tag{2}$$

$$\sigma = e^2 \int \Xi(\varepsilon)\left(-\frac{\partial f_0}{\partial \varepsilon}\right) d\varepsilon, \tag{3}$$

$$\kappa_e = k_B^2 T \int \Xi(\varepsilon)\left(-\frac{\partial f_0}{\partial \varepsilon}\right)\left[\frac{\varepsilon - \mu}{k_B T}\right]^2 d\varepsilon - TS^2\sigma. \tag{4}$$

Here, $e$, $f_0$ and $\mu$ are the electron charge, Fermi distribution function and chemical potential, respectively. The transport distribution $\Xi = \sum_{\mathbf{k}} v_{\mathbf{k}} v_{\mathbf{k}} \tau_{\mathbf{k}}$ with $v_{\mathbf{k}}$ the group velocity and $\tau_{\mathbf{k}}$ the relaxation time at state $\mathbf{k}$. The last term $TS^2\sigma$ in Eq. (4) is usually ignored due to its small effect when the power factor and the temperature are low [31]. At elevated temperature, however, we should explicitly consider it. The relaxation time can be calculated by using deformation potential theory [32] assuming acoustic phonons are the main scattering mechanism, as given by

$$\tau = \frac{2\sqrt{2\pi} C \hbar^4}{3(k_B T m_{dos}^*)^{3/2} E^2}, \tag{5}$$

where $k_B$, $C$, $m_{dos}^*$ and $E$ are the Boltzmann constant, elastic constant, DOS effective mass and deformation potential constant, respectively. Here, $m_{dos}^*$ can be obtained by $m_{dos}^* = N^{2/3}(m_x^* m_y^* m_z^*)^{1/3}$ [33] with $N$ the valley degeneracy. $m_x^*$, $m_y^*$ and $m_z^*$ are the band effective masses along $x$, $y$ and $z$ directions, respectively. All these parameters are summarized in Table II. In Figure 3, we plot the relaxation time as a function of temperature, where we see $\tau$ of $n$-type carrier along the $x$ and $z$ directions are almost identical to each other and are obviously larger than those of $p$-type carrier. The major reason is that $m_{dos}^*$ of electron is significantly smaller than that of hole as characterized by the different band curvature and multi-valley structures shown in Figure 2. At room temperature, the relaxation time of $n$-type KSnSb can reach as high as 40 fs, which is larger than those of good thermoelectric materials such as $Bi_2Te_3$ (~22 fs) [34] and $Sb_2Te_3$ (~20 fs) [6] and is very beneficial for its thermoelectric performance.

Figure 4 shows the calculated Seebeck coefficient (absolute value), the electrical conductivity, and the power factor of KSnSb as a function of carrier concentration at two typical temperatures of 300 K and 800 K. We do not consider higher temperature since it was previously found that the melting point of KSnSb should be lower than the temperature of synthesis reaction (803 K) [12]. As can be seen from Figure 4(a) and 4(c), the absolute values of the Seebeck coefficient |S| decrease with the carrier concentration, and are almost identical to each other along the *x* and *z* directions. At room temperature, the absolute values of Seebeck coefficients can be reached to ~250 μV/K at carrier concentration of $1 \times 10^{19}$ cm$^{-3}$, which is even higher than that of SnSe (~200 μV/K) [35]. When the temperature is increased to 800 K, the Seebeck coefficients become even larger. The origin of such high Seebeck coefficient should be attributed to the multi-valley band structures. As indicated in Figure 2, there are two energy valleys around the CBM and VBM, which lead to a significant increase of the DOS $g(\varepsilon)$ around the Fermi level and enhanced energy-dependence of the carrier density $n(\varepsilon)$. According to the Mott relation [36]

$$S = \frac{\pi^2 k_B^2 T}{3q} \left\{ \frac{1}{n} \frac{dn(\varepsilon)}{d\varepsilon} + \frac{1}{\mu_c} \frac{d\mu_c(\varepsilon)}{d\varepsilon} \right\}, \qquad (6)$$

where $q$ is the carrier charge and $\mu_c$ is the carrier mobility. It is obvious that enhanced $\frac{dn(\varepsilon)}{d\varepsilon}$ will results in higher Seebeck coefficient.

As opposite to the Seebeck coefficient, the electrical conductivity shown in Figure 4(a) and 4(c) increases with carrier concentration. Moreover, we note that the electrical conductivity of *n*-type system is several orders of magnitude larger than that of *p*-type, which is also found in the state-of-the-art thermoelectric material SnSe [35]. Such observation is due to the fact that the mobility of electron is much higher than that of hole caused by smaller effective mass. Since both *p*- and *n*-type systems exhibit similar Seebeck coefficients, the power factor of *n*-type system is larger than that of *p*-type, as indicated in Figure 4(b) and 4(d). According to the Wiedemann–Franz law [37], the electronic thermal conductivity $\kappa_e$ is connected with the electrical conductivity $\sigma$ by $\kappa_e = L\sigma T$, where $L$ is the Lorenz number. In principle, the variation of $\kappa_e$ with carrier concentration is almost the same as that of $\sigma$ and is thus not shown here.

**3.2 Phonon transport properties**

We now move to the discussion of phonon transport in KSnSb. The corresponding thermal conductivity $\kappa_p$ can be figured out according to the phonon Boltzmann theory [25]

$$\kappa_p = \frac{1}{N_0 \Omega k_B T^2} \sum_\lambda n_\lambda^0 \left(n_\lambda^0 + 1\right)\left(v_\lambda\right)^2 \tau_\lambda \omega_\lambda^2, \tag{7}$$

where $N_0$, $\Omega$, $n_\lambda^0$, $v_\lambda$, $\tau_\lambda$ and $\omega_\lambda$ are the number of ***q***-points in the first Brillouin zone, the volume of the primitive cell, the Bose-Einstein distribution function, the group velocity, the relaxation time, and the angular frequency of phonon mode $\lambda$, respectively. As shown in Figure 5, the $\kappa_p$ decreases monotonously with temperature increasing from 300 to 800 K. Moreover, the $\kappa_p$ along the *z* direction (out-of-plane) is obviously lower than that along the *x* direction (in-plane), which is generally found in many layered thermoelectric materials such as $Bi_2Te_3$ [34], $Sb_2Te_3$ [6] and SnSe [38]. On the other hand, we find that the lattice thermal conductivity of KSnSb is rather small. For example, the *x* and *z* component at 300 K are 1.6 and 0.9 W/mK, respectively. To understand the origin of such intrinsically lower thermal conductivity, we first investigate the phonon dispersion relations of KSnSb, as shown in Figure 6(a). We see there is no imaginary frequency which indicates that the structure is dynamically stable. Note that the maximum phonon frequency of KSnSb is about 5.3 THz. Such lower value is comparable with those in $Bi_2Te_3$ (4.6 THz) [34], $Sb_2Te_3$ (5.2 THz) [39], and SnSe (5.6 THz) [35]. Besides, there are two large phonon gaps around 2 and 4 THz, which is a sign of low lattice thermal conductivity [40]. The analysis of phonon density of states (PDOS) indicates that the phonon branches in the lower and higher frequency region are mainly contributed by the Sn and Sb atoms, while the K atoms dominate the intermediate frequency region.

In Figure 6(b), we show the lattice thermal conductivity contribution as a function of frequency along the *x* and *z* directions at 300 K. It can be found that more than 90% of $\kappa_p$ is contributed by low-frequency phonons (< 2 THz), or caused by the Sn and Sb atoms as discussed above. In addition, we find from Figure 6(c) that the average group velocity $v_\lambda$ below 2 THz is only about 2.0 km/s, which is similar to that of the thermoelectric material PbTe [41]. Such small value again suggests the low lattice thermal conductivity of KSnSb.

The normalized cumulative lattice thermal conductivity of KSnSb with respect to phonon mean free paths (MFP) at 300 K is shown in Figure 6(d). It is clear that $\kappa_p$ is dominated by phonons with MFP ranging from 1 to 200 nm. Besides, the MFP values for 50% $\kappa_p$ accumulation are 23 and 19 nm for the *x* and *z* directions, respectively. Such lower MFP values at half $\kappa_p$ are close to that found in PbTe (~20 nm) [42]. Meanwhile, the mean phonon scattering rate of KSnSb we obtained is 0.2 ps$^{-1}$ which is similar to that of SnSe [38]. The calculated total Grüneisen parameter $\gamma$ of KSnSb is about 1.0, which is ~28% lower than that of Bi$_2$Te$_3$ ($\gamma = 1.4$) [34] and ~59% higher than that of SnSe ($\gamma = 0.63$) [38]. The anharmonic phase space volume $P_3$ of KSnSb is calculated to be $1.3 \times 10^{-2}$ eV, which is a bit lower than that of SnSe ($1.9 \times 10^{-2}$ eV) [38] and much higher than that of Si ($0.35 \times 10^{-2}$ eV) [43]. All these findings confirms that the layered KSnSb indeed have intrinsically lower lattice thermal conductivity, and is in principle governed by the co-existence of K-K metallic bonds, K-Sn ionic bonds, Sn-Sb covalent bonds, as well as the interlayer vdW interactions [44].

**3.3 Figure of merit**

After all the transport coefficients have been calculated, we can now predict the thermoelectric performance of KSnSb. Figure 7 shows the *ZT* values as a function of temperatures. As *n*-type system exhibits much higher power factor than *p*-type system (Figure 4), we see that the *n*-type *ZT* is significantly larger than *p*-type one. Besides, the *ZT* value of *n*-type KSnSb shows strong temperature dependence, which is however less for the *p*-type system. In Table III, we summarize the *ZT* values of *n*-type KSnSb and the corresponding transport coefficients from 300 K to 800 K. At optimized carrier concentration of $3.2 \times 10^{19}$ cm$^{-3}$ and $3.3 \times 10^{19}$ cm$^{-3}$, the maximum *ZT* values of 2.6 and 2.4 can be reached at 800 K along the *x* and *z* directions, respectively. Such extraordinary thermoelectric performance of KSnSb exceeds those of most bulk thermoelectric materials. Moreover, we see that the in-plane *ZT* value is comparable to that of out-of-plane, which suggests that the polycrystalline KSnSb may exhibit almost the same electronic transport properties as single crystal. Consider the fact that polycrystalline could have smaller lattice thermal conductivity, it is expected that polycrystalline KSnSb may exhibit even better thermoelectric performance.

## 4. Summary

In summary, we demonstrate by first-principles calculations that the Zintl compound KSnSb is semiconducting with an indirect band gap of 0.75 eV predicted from hybrid functional. The multi-valley band structures around the Fermi level lead to significantly enhanced Seebeck coefficient and power factor, as derived from the Boltzmann transport theory with carrier relaxation time obtained by the deformation theory. By solving the phonon Boltzmann transport equation, we find that the KSnSb compound exhibits very small lattice thermal conductivity, as characterized by its large values of phonon gaps, Grüneisen parameter, and phase space volume, as well as small values of group velocity and phonon mean free path. In principle, the intrinsically lower lattice thermal conductivity is traced back to the mixture of various bonding in the KSnSb compound, including the K-K metallic bonds, the K-Sn ionic bonds, the Sn-Sb covalent bonds, and the interlayer vdW interactions. All these findings make KSnSb a very promising candidate for thermoelectric application in the high temperature region, with a record high *ZT* value of 2.6 at optimized electron concentration of $3.2\times10^{19}$ cm$^{-3}$.

**Acknowledgments**

We thank financial support from the National Natural Science Foundation (Grant No. 11574236 and 51172167) and the "973 Program" of China (Grant No. 2013CB632502).

**Table I** The calculated structural parameters of Zintl compound KSnSb with and without consideration of vdW interactions. The experimental data of Ref. [12] are also shown for comparison.

| This work | Exp. | w/o vdW | with vdW |
|---|---|---|---|
| $a$ (Å) | 4.359 | 4.45 | 4.40 |
| $c$ (Å) | 13.15 | 13.29 | 13.00 |
| $z_{Sn}$ | 0.2010 | 0.203 | 0.200 |
| $z_{Sb}$ | 0.3115 | 0.312 | 0.312 |

**Table II** The density of state effective mass $m^*_{dos}$, the elastic constant $C$, and the deformation potential constant $E$ of $n$- and $p$-type KSnSb along the $x$ and $z$ directions.

| Carrier type | $m^*_{dos}$ ($m_e$) | $C_x$ (Gpa) | $E_x$ (eV) | $C_z$ (Gpa) | $E_z$ (eV) |
|---|---|---|---|---|---|
| electron | 0.266 | 55.2 | 9.21 | 44.3 | 8.37 |
| hole | 1.67 | 55.2 | 6.18 | 44.3 | 10.4 |

**Table III** The optimized *ZT* values of *n*-type KSnSb along the *x* and *z* directions at 300 and 800 K. The corresponding carrier concentration and transport coefficients are also listed.

|   | $T$ (K) | $n$ ($10^{19}$cm$^{-3}$) | $S$ (μV/K) | $\sigma$ (S/cm) | $S^2\sigma$ ($10^{-3}$W/mK$^2$) | $\kappa_e$ (W/mK) | $\kappa_p$ (W/mK) | $ZT$ |
|---|---|---|---|---|---|---|---|---|
| *x* | 300 | 3.0 | −188 | 1773 | 6.24 | 1.24 | 1.61 | 0.7 |
|   | 400 | 2.9 | −227 | 1103 | 5.69 | 1.00 | 1.20 | 1.0 |
|   | 500 | 2.9 | −260 | 762 | 5.14 | 0.83 | 0.96 | 1.4 |
|   | 600 | 3.0 | −286 | 573 | 4.68 | 0.72 | 0.80 | 1.8 |
|   | 700 | 3.1 | −307 | 454 | 4.27 | 0.64 | 0.69 | 2.3 |
|   | 800 | 3.2 | −323 | 380 | 3.96 | 0.60 | 0.60 | 2.6 |
| *z* | 300 | 3.0 | −179 | 949 | 3.04 | 0.64 | 0.92 | 0.6 |
|   | 400 | 3.0 | −219 | 563 | 2.69 | 0.50 | 0.69 | 0.9 |
|   | 500 | 2.8 | −250 | 392 | 2.45 | 0.43 | 0.55 | 1.3 |
|   | 600 | 2.9 | −276 | 297 | 2.26 | 0.38 | 0.46 | 1.6 |
|   | 700 | 3.1 | −296 | 242 | 2.11 | 0.35 | 0.39 | 2.0 |
|   | 800 | 3.3 | −312 | 203 | 1.98 | 0.33 | 0.34 | 2.4 |

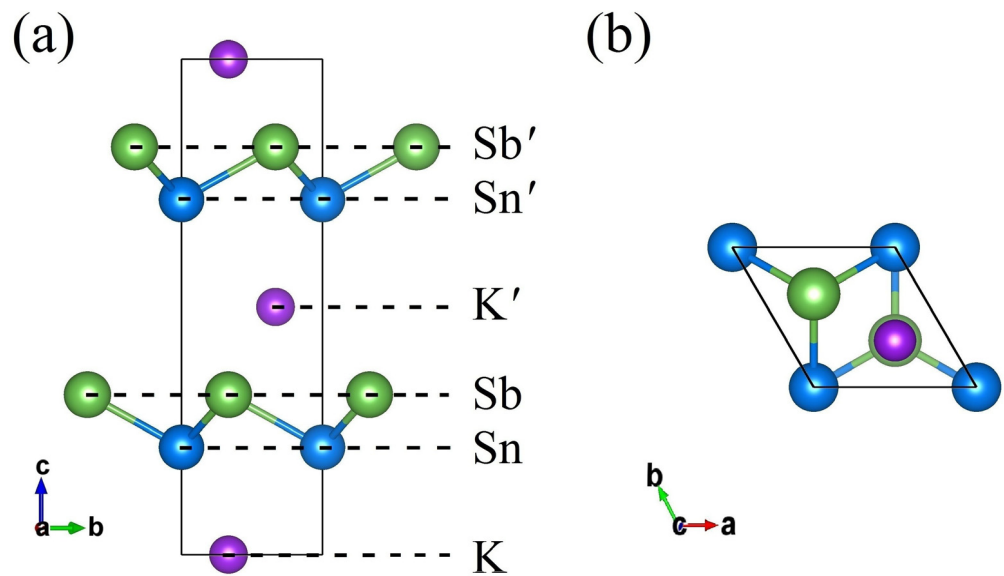

**Figure 1** The crystal structure of KSnSb (a) side-view, and (b) top-view.

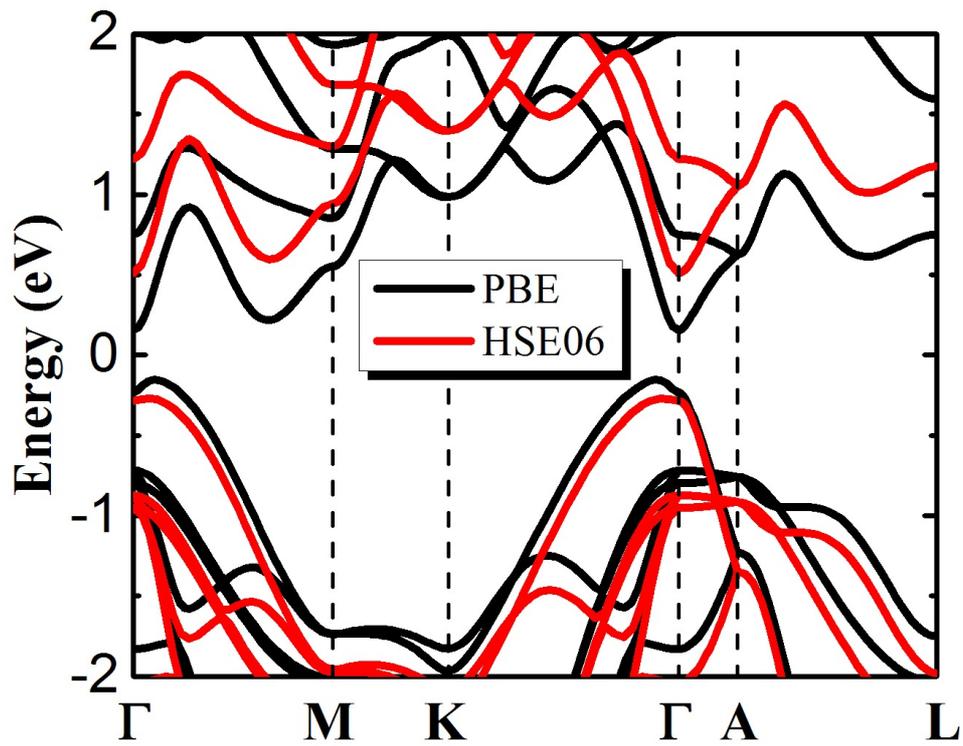

**Figure 2** The energy band structures of KSnSb, calculated with PBE and HSE06 functionals. The Fermi level is at 0 eV.

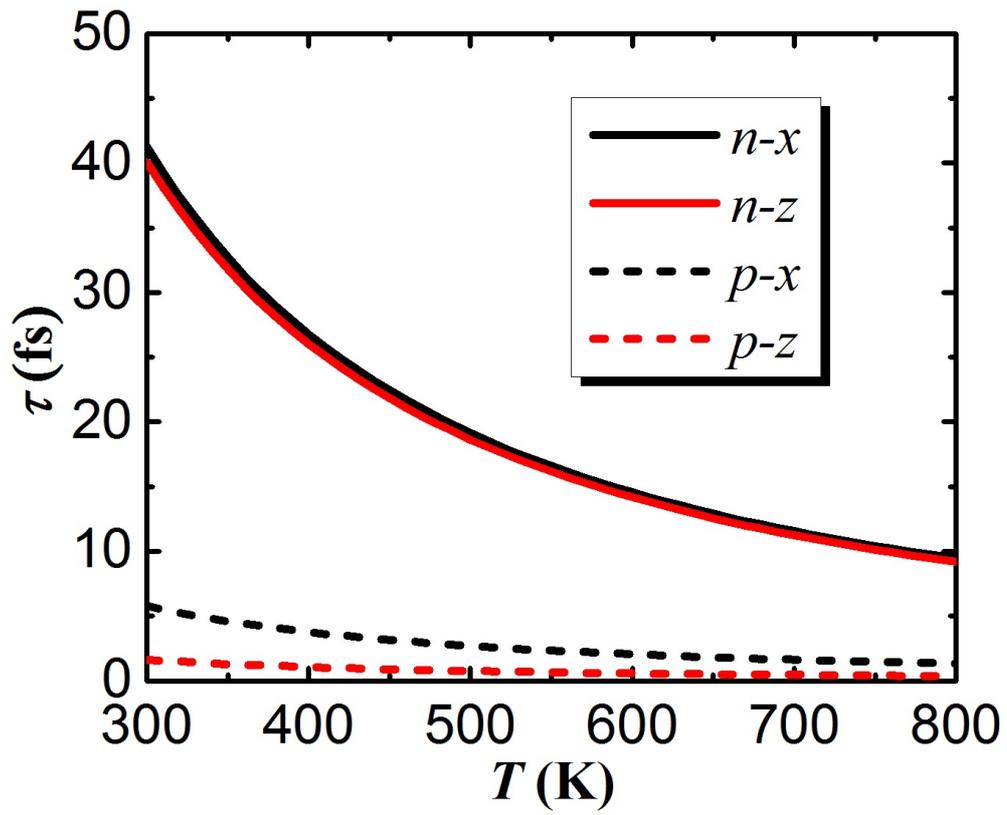

**Figure 3** The relaxation time as a function of temperature for KSnSb along the $x$ (in-plane) and $z$ (out-of-plane) directions.

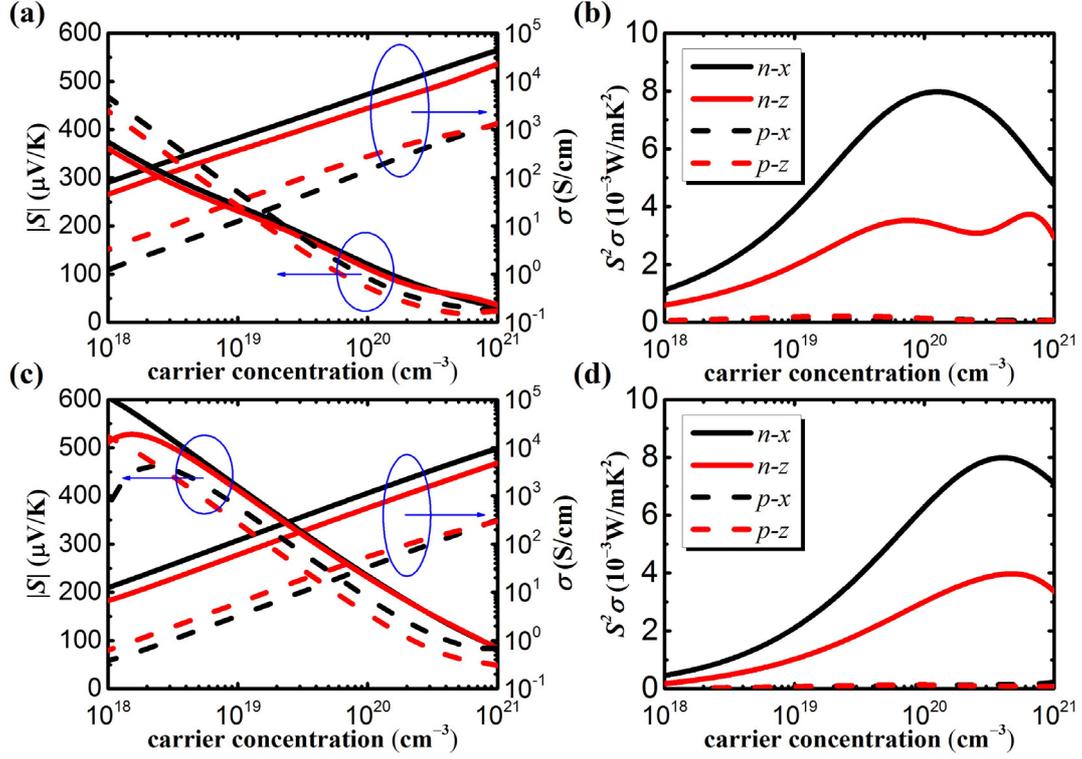

**Figure 4** The calculated electronic transport coefficients as a function of carrier concentration along the $x$ and $z$ directions for $n$-type (solid line) and $p$-type KSnSb (dashed line): (a) the Seebeck coefficient (absolute value) $|S|$ and the electrical conductivity $\sigma$ at 300 K, (b) the power factor $S^2\sigma$ at 300 K, (c) the Seebeck coefficient (absolute value) $|S|$ and the electrical conductivity $\sigma$ at 800 K, (d) the power factor $S^2\sigma$ at 800 K.

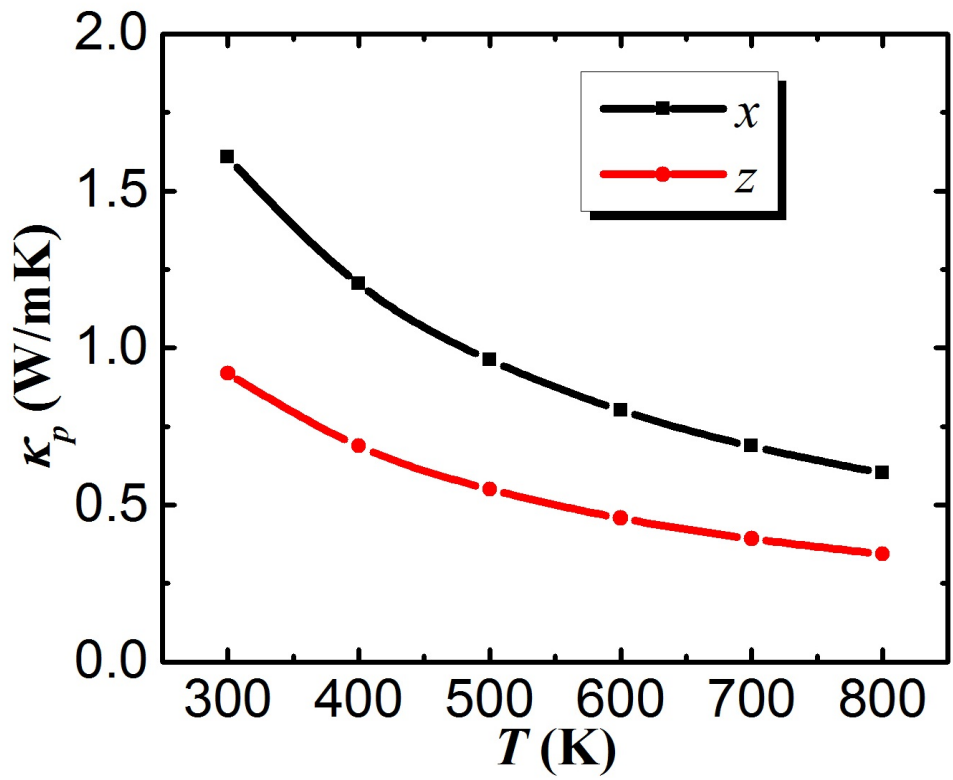

**Figure 5** The lattice thermal conductivity as a function of temperature for KSnSb along the $x$ (in-plane) and $z$ (out-of-plane) directions.

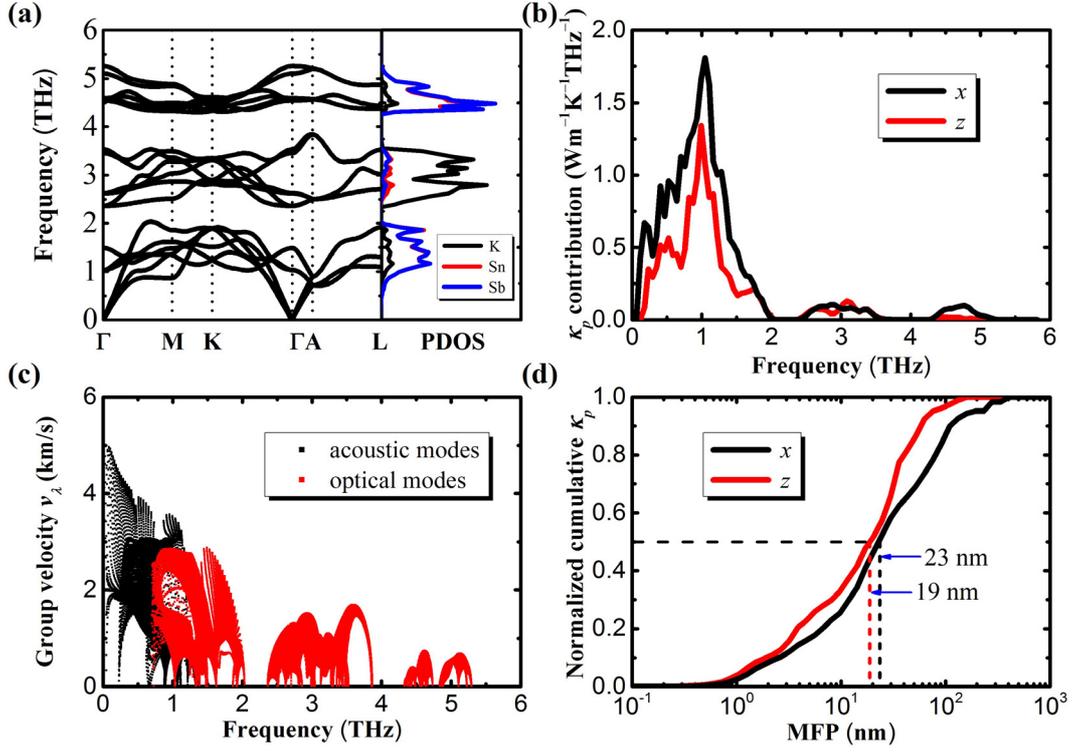

**Figure 6** (a) The phonon dispersion relations and phonon density of states for KSnSb, (b) the lattice thermal conductivity contribution as a function of phonon frequency along the *x* and *z* directions at 300 K, (c) the phonon group velocity as a function of phonon frequency, and (d) the normalized cumulative lattice thermal conductivity as a function of phonon mean free path.

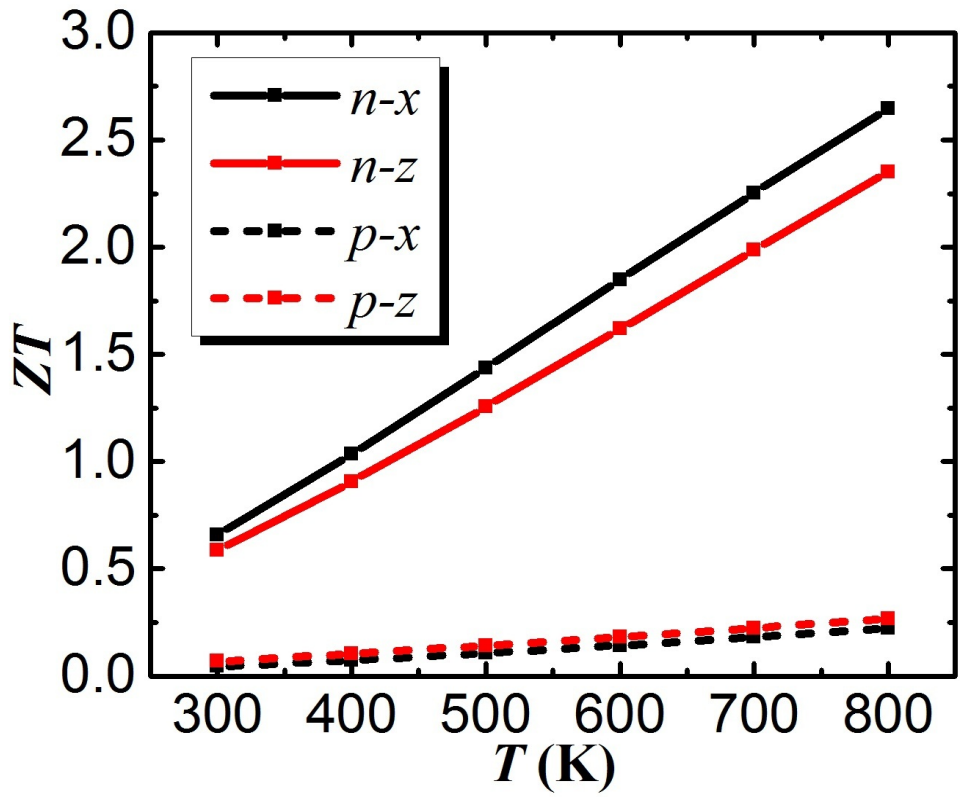

**Figure 7** The calculated *ZT* values of KSnSb as a function of temperature.

# References

[1] G. A. Slack, in *CRC Handbook of Thermoelectrics*, edited by D. M. Rowe (CRC Press, London, 1995), p. 407.

[2] S. M. Kauzlarich, S. R. Brown, and G. J. Snyder, Dalton Transactions **21**, 2099 (2007).

[3] Y. Hu, J. Wang, A. Kawamura, K. Kovnir, and S. M. Kauzlarich, Chem. Mater. **27**, 343 (2015).

[4] T. Yamada, H. Yamane, and H. Nagai, Adv. Mater. **27**, 4708 (2015).

[5] J. Shuai, H. Geng, Y. Lan, Z. Zhu, C. Wang, Z. Liu, J. Bao, C. W. Chu, J. Sui, and Z. Ren, PNAS **113**, E4125 (2016).

[6] J. Zhou, Y. Wang, J. Sharp, and R. Yang, Phys. Rev. B **85**, 115320 (2012).

[7] M. Liu, X. Qin, C. Liu, L. Pan, and H. Xin, Phys. Rev. B **81**, 245215 (2010).

[8] A. Möchel, I. Sergueev, H. C. Wille, F. Juranyi, H. Schober, W. Schweika, S. R. Brown, S. M. Kauzlarich, and R. P. Hermann, Phys. Rev. B **84**, 184303 (2011).

[9] P. Alemany, M. Llunell, and E. Canadell, J. Comput. Chem. **29**, 2144 (2008).

[10] P. Gorai, E. S. Toberer, and V. Stevanović, J. Mater. Chem. A **4**, 11110 (2016).

[11] L. D. Zhao, S. H. Lo, Y. Zhang, H. Sun, G. Tan, C. Uher, C. Wolverton, V. P. Dravid, and M. G. Kanatzidis, Nature **508**, 373 (2014).

[12] K. H. Lii and R. C. Haushalter, J. Solid State Chem. **67**, 374 (1987).

[13] P. C. Schmidt, D. Stahl, B. Eisenmann, R. Kniep, V. Eyert, and J. Kübler, J. Solid State Chem. **97**, 93 (1992).

[14] P. Alemany, M. Llunell, and E. Canadell, J. Comput. Chem. **29**, 2144 (2008).

[15] G. Kresse and J. Hafner, Phys. Rev. B **47**, 558 (1993).

[16] G. Kresse and J. Hafner, Phys. Rev. B **49**, 14251 (1994).

[17] G. Kresse and J. Furthmüller, Comput. Mater. Sci. **6**, 15 (1996).

[18] J. P. Perdew, K. Burke, and M. Ernzerhof, Phys. Rev. Lett. **77**, 3865 (1996).

[19] J. Heyd, G. E. Scuseria, and M. Ernzerhof, J. Chem. Phys. **118**, 8207 (2003).

[20] J. Heyd and G. E. Scuseria, J. Chem. Phys. **121**, 1187 (2004).

[21] J. Heyd, G. E. Scuseria, and M. Ernzerhof, J. Chem. Phys. **124**, 219906 (2006).

[22] T. Thonhauser, V. R. Cooper, S. Li, A. Puzder, P. Hyldgaard, and D. C. Langreth, Phys. Rev. B **76**, 125112 (2007).

[23] T. J. Scheidemantel, C. Ambrosch-Draxl, T. Thonhauser, J. V. Badding, and J. O. Sofo, Phys. Rev. B **68**, 125210 (2003).

[24] P. Hohenberg and W. Kohn, Phys. Rev. **136**, B864 (1964).

[25] W. Li, J. Carrete, N. A. Katcho, and N. Mingo, Comput. Phys. Commun. **185**, 1747 (2014).

[26] A. Togo and I. Tanaka, Scripta Mater. **108**, 1 (2015).

[27] W. Li, L. Lindsay, D. A. Broido, D. A. Stewart, and N. Mingo, Phys. Rev. B **86**, 174307 (2012).

[28] J. Mao, W. Liu, and Z. Ren, Journal of Materiomics **2**, 203 (2016).

[29] G. Wang and T. Cagin, Phys. Rev. B **76**, 075201 (2007).

[30] L. D. Zhao, G. J. Tan, S. Q. Hao, J. Q. He, Y. L. Pei, H. Chi, H. Wang, S. K. Gong, H. B. Xu, V. P. Dravid, C. Uher, G. J. Snyder, C. Wolverton, and M. G. Kanatzidis, Science **351**, 141 (2016).

[31] G. K. H. Madsen and D. J. Singh, Comput. Phys. Commun. 175, 67 (2006).

[32] J. Bardeen and W. Shockley, Phys. Rev. **80**, 72 (1950).

[33] Y. Pei, X. Shi, A. Lalonde, H. Wang, L. Chen, and G. J. Snyder, Nature **473**, 66 (2011).

[34] B. L. Huang and M. Kaviany, Phys. Rev. B 77, 125209 (2008).

[35] R. Guo, X. Wang, Y. Kuang, and B. Huang, Phys. Rev. B **92**, 115202 (2015).

[36] J. P. Heremans, V. Jovovic, E. S. Toberer, A. Saramat, K. Kurosaki, A. Charoenphakdee, S. Yamanaka, and G. J. Snyder, Science **321**, 554 (2008).

[37] C. Kittel, Am. J. Phys. **35**, 547 (1967).

[38] J. Carrete, N. Mingo, and S. Curtarolo, Appl. Phys. Lett. **105**, 101907 (2014).

[39] N. A. Katcho, N. Mingo, and D. A. Broido, Phys. Rev. B **85**, 035436 (2012).

[40] B. Peng, H. Zhang, H. Shao, Y. Xu, G. Ni, R. Zhang, and H. Zhu, Phys. Rev. B **94**, 245420 (2016).

[41] Z. Tian, J. Garg, K. Esfarjani, T. Shiga, J. Shiomi, and G. Chen, Phys. Rev. B **85**, 184303 (2012).

[42] B. Qiu, H. Bao, G. Zhang, Y. Wu, and X. Ruan, Comput. Mater. Sci. **53**, 278 (2012)

[43] L. Lindsay and D. A. Broido, J. Phys.: Condens. Matter **20**, 165209 (2008).

[44] D. Yang, W. Yao, Q. Chen, K. Peng, P. Jiang, X. Lu, C. Uher, T. Yang, G. Wang, and X. Zhou, Chem. Mater. **28**, 1611 (2016).